\documentclass{jpp}
\usepackage{graphicx}

\usepackage[utf8]{inputenc}
\usepackage[T1]{fontenc}
\usepackage{amsmath}
\usepackage{amssymb}

\usepackage{here}
\usepackage[usenames,dvipsnames]{xcolor}
\usepackage{url}
\usepackage[colorlinks=true,
		linkcolor=red,
		citecolor=blue,
		urlcolor=blue]{hyperref}
\usepackage{tabularx,ragged2e,booktabs}
\usepackage{bm}
\usepackage{rotating}
\usepackage{mathtools}
\usepackage{epstopdf}
\newcommand{\ordo}{\textit{O}}
\newcommand{\be}{\begin{equation}} \newcommand{\ee}{\end{equation}}

\renewcommand{\thesection}{\Roman{section}}
\renewcommand{\thesubsection}{\Alph{subsection}}

\newcommand{\vpa}{v_{\|}}
\newcommand{\vpe}{v_{\perp}}

\newcommand{\xgc}{{\sc XGC}}
\newcommand{\petsc}{{\sc pets}c}

\newcommand{\overbar}[1]{\mkern 1.5mu\overline{\mkern-1.5mu#1\mkern-1.5mu}\mkern 1.5mu}

\newcommand{\longdashed}{\protect\mbox{\rule[1.0pt]{0.8cm}{3.0pt} \; \rule[1.0pt]{0.8cm}{3.0pt}}}

\newcommand{\dashdot}{\protect\mbox{\rule[1.0pt]{0.6cm}{3.0pt} \; \rule[1.0pt]{3.0pt}{3.0pt} \; \rule[1.0pt]{0.6cm}{3.0pt}}}

\newcommand{\newfull}{\protect\mbox{\rule[1.0pt]{2.0cm}{3.0pt}}}

\definecolor{ColorZeff1p05Two}{HTML}{BBBB00}
\definecolor{ColorZeff1p05Four}{HTML}{444422}
\definecolor{ColorZeff1p05Five}{HTML}{BB0000}
\definecolor{ColorZeff1p05Six}{HTML}{FF9900}

\definecolor{ColorZeff2p0Three}{gray}{0.7}
\definecolor{ColorZeff2p0Four}{HTML}{CCCCAA}
\definecolor{ColorZeff2p0Five}{HTML}{22FF22}

\definecolor{NewRed}{HTML}{FF0000}
\definecolor{NewBlue}{HTML}{1F77B4}
\definecolor{NewCyan}{HTML}{00AAAA}
\definecolor{NewMagenta}{HTML}{AA00AA}
\definecolor{NewGreen}{HTML}{2CA02C}
\definecolor{NewYellow}{HTML}{AAAA00}
\definecolor{NewOrange}{HTML}{FF7F0E}

\shorttitle{Implementation of higher-order velocity mapping}
\shortauthor{A.~Moll\'en, M.~F.~Adams, M.~G.~Knepley, R.~Hager and C.~S.~Chang}

\title{Implementation of higher-order velocity mapping between marker particles and grid in the particle-in-cell code XGC}

\author{Albert~Moll\'en\aff{1}
  \corresp{\email{\href{mailto:amollen@pppl.gov}{amollen@pppl.gov}}},
  M.~F.~Adams\aff{2},
  M.~G.~Knepley\aff{3},
  R.~Hager\aff{1}
 \and C.~S.~Chang\aff{1}}

\affiliation{\aff{1}Princeton Plasma Physics Laboratory, Princeton, NJ, USA
\aff{2}Lawrence Berkeley National Laboratory, Berkeley, CA, USA
\aff{3}State University of New York at Buffalo, NY, USA}

\begin{document}

\maketitle

\begin{abstract} 
The global total-$f$ gyrokinetic particle-in-cell code \xgc, used to study transport in magnetic fusion plasmas, implements a continuum grid to perform the dissipative operations, such as plasma collisions. To transfer the distribution function between marker particles and a rectangular velocity-space grid, \xgc~employs a bilinear mapping. 
The conservation of particle density and momentum is accurate enough in this bilinear operation, but the error in the particle energy conservation can become undesirably large in special conditions. 
In the present work we update \xgc~to use a novel mapping technique, based on the calculation of a pseudo-inverse, to exactly preserve moments up to the order of the discretization space. We describe the details of the implementation and we demonstrate the reduced interpolation error for a neoclassical tokamak test case by using $1^{\mathrm{st}}$- and $2^{\mathrm{nd}}$-order elements with the pseudo-inverse method and comparing to the bilinear mapping.
\end{abstract}

\section{Introduction}\label{sec:Introduction}
The X-point Gyrokinetic Code (\xgc) is a total-$f$ gyrokinetic particle-in-cell (PIC) code used to study kinetic transport phenomena, optimized for solving the neoclassical and the turbulent dynamics together with the neutral particle transport, in magnetic fusion plasmas \citep{KuPoP2018}. \xgc~is a global code treating the whole closed-field-line portion of a plasma, together with the open-field-line region, referred to as the ``scrape-off layer (SOL)'' across the magnetic separatrix surface. The evolution of plasma species $s$ is described by the particle distribution function $\displaystyle f_s\left(\boldsymbol{X}, \vpa, \vpe, t\right)$, where $\boldsymbol{X}$ is the 3D ``gyrocenter'' position, $\displaystyle \vpa = \boldsymbol{v} \cdot \boldsymbol{B} / \left|\boldsymbol{B}\right|$ and $\vpe = \sqrt{v^2 - \vpa^2}$ the velocity parallel and perpendicular to the magnetic field $\boldsymbol{B}$ respectively, and $t$ represents time. The distribution functions are obtained by solving the 5D gyrokinetic-Maxwell system of equations, and from the functions quantities of interest, such as fluxes, can be derived. 
In addition to PIC methods, \xgc~uses Eulerian/continuum methods for dissipative operations such as collisions and heat sources. This requires a mapping between the marker particles and a 5D phase-space grid. In velocity space \xgc~employs a bilinear mapping, which inherently lacks energy conservation. Although the energy error can be controlled by the number of marker particles \citep[see][]{HagerJCP2016}, a quadratic $2^{\mathrm{nd}}$-order mapping (or even higher order) is desirable because it enables exact energy conservation. This work describes the implementation of a $2^{\mathrm{nd}}$-order mapping technique that exactly conserves mass, momentum and energy.

In the code the total distribution function is divided as 
\begin{equation}
f_s = f_0 + \delta f = f_a + f_g + f_p,
\label{eq:XGCtotalDistribution}
\end{equation}
where $\delta f = f_p$ is represented by marker particles carrying the fast varying physics information such as particle orbital and turbulent dynamics, and $f_0$ is split into a slowly varying part, $f_g$, represented on a coarse 5D phase-space continuum grid, and an analytic function, $f_a$, representing the plasma Maxwellian background.  Both $f_g$ and $f_p$ can be highly non-Maxwellian, taking the whole $f_s$ to the highly non-equilibrium state. $f_a$ can also be updated from $f_g$ and $f_p$.
The 5D grid consists of a 3D unstructured triangular mesh in configuration space and a 2D structured rectangular grid in velocity space. 
\xgc~can operate with different versions of the distribution function in Eq.~\eqref{eq:XGCtotalDistribution} \citep[for details see discussion about ``full-f'', ``reduced $\delta f$'' and ``total-$\delta f$'' mode in][]{KuPoP2018}, and the present work mainly focuses on the ``total-$\delta f$'' scheme where fractions of $f_p$ can be transferred over to $f_g$ and the Maxwellian part of $f_g$ to $f_a$ during a time step. 
To evaluate the dissipative operations, such as plasma collisions, \xgc~presents the full distribution function on the 5D grid \citep{HagerJCP2016,YoonPoP2014}. 
\xgc~implements a multi-species nonlinear Fokker-Planck-Landau collision operator \citep{Landau1936}. 
The choice of implementing the Coulomb collision operation using a mesh-based numerical scheme rather than a particle-particle based scheme, has to do with the fact that the computational cost of a particle-particle scheme is at least $\sim \ordo\left(N_p\right)$ per species where $N_p$ is the number of marker particles, whereas the cost of a mesh-based scheme is $\sim \ordo\left(N_g\right)$ where $N_g$ is the number of grid points. For a typical simulation the required grid resolution has $N_g \ll N_p$, and the cost of mapping the particles back-and-forth between continuous particle position and the 5D grid is small (although theoretically $\sim \ordo\left(N_p\right)$) making a mesh-based scheme much cheaper \citep{YoonPoP2014}. The 5D grid is used not only for Coulomb collisions, but for other dissipative operations such as heat sources and sinks, and plasma interaction with neutral particles.
An additional advantage of the mesh-based scheme is that it can be used to exchange the particle distribution function between \xgc~and coupled continuum codes, as is done in the ``High-Fidelity Whole Device Modeling of Magnetically Confined Fusion Plasma'' project of the Exascale Computing Program \citep{DominskiPoP2018,DominskiPoP2020}.

The present mapping scheme between marker particles and grid in 2D velocity space utilized in \xgc~is described in \citep[Sec.~III]{YoonPoP2014}. 
However, \citep{YoonPoP2014} discusses the implementation in an older ``full-f'' version of \xgc~and the newer ``total-$\delta f$'' version contains a minor modification in the inverse mapping which will be detailed later in Sec.~\ref{sec:LagrangeElements}.
The mapping is implemented using bilinear shape functions to map the marker particle weights onto the uniform rectangular grid and an inverse mapping back to the particles. 
This bilinear interpolation conserves particle number and momentum to high enough accuracy but fails to conserve energy to a desired degree in special situations that are often targeted by \xgc. 
This implies that if the dissipative operations are applied on the velocity grid the energy conservation chain is destroyed
even if the Coulomb collision operation itself conserves particle number, parallel momentum and energy. 
The inevitable interpolation error is a consequence of the fact that a discretization space can only preserve moments up to the order of the space, i.e. $1^{\mathrm{st}}$-order moments for a linear mapping. To ensure energy conservation we have to implement $2^{\mathrm{nd}}$-order elements or above. 

The purpose of this work is to update \xgc~to use a novel mapping technique, recently included in the \petsc~library \citep{petsc2020}, which employs a pseudo-inverse to preserve moments up to the order of the discretization space \citep{HirvijokiArxiv2018}. 
We demonstrate the effectiveness of this technique by comparing the interpolation error in particle number, parallel momentum and kinetic energy when using $2^{\mathrm{nd}}$-order elements to the former bilinear mapping in a neoclassical \xgc~simulation. 

The remainder of the paper is organized as follows. 
In Sec.~\ref{sec:LagrangeElements}, we discuss the current velocity interpolation in \xgc~based on $1^{\mathrm{st}}$-order Lagrange elements and show the numerical complication of extending this to $2^{\mathrm{nd}}$-order elements to preserve energy. 
Sec.~\ref{sec:NovelTechnique} introduces a novel technique as a solution to overcome the numerical difficulties. 
We then use the new scheme in \xgc~to conduct neoclassical simulations for a circular tokamak test case and show the improved conservation properties in Sec.~\ref{sec:Application}. 
Finally, we conclude in Sec.~\ref{sec:conclusions} and discuss future developments in \xgc. 
%
\section{Lagrange elements}\label{sec:LagrangeElements}
The particle distribution function $f_p$ in Eq.~\eqref{eq:XGCtotalDistribution} is represented by marker particles $k$ with weights $w_{1,k}$, $w_{0,k}$ and continuous 5D phase-space positions $\boldsymbol{z}_k = \left(\boldsymbol{X}_k, v_{\| k}, v_{\perp k}\right)$,
\begin{equation}
f_p \left(\boldsymbol{z}\right) = \sum_{k} w_{1,k} \left. f_0\left(\boldsymbol{z}_k, t\right)\right|_{t = 0} \; \delta\left(\boldsymbol{z} - {\boldsymbol{z}_k}\right) = \sum_{k} w_{1,k} \, w_{0,k} \, g \; \delta\left(\boldsymbol{z} - {\boldsymbol{z}_k}\right) .
\label{eq:deltafDistribution}
\end{equation}
Here $\displaystyle \delta\left(\boldsymbol{z}\right)$ is the Dirac delta function. 
\xgc~uses a two-weight scheme: a full\nobreakdash-$f$ weight $w_{0,k}$ and a delta\nobreakdash-$f$ weight $w_{1,k}$ for each marker particle \citep{KuPoP2018,KuJCP2016}. 
$w_{1,k} \, w_{0,k}$ effectively is the number of real particles represented by the marker particle. 
For the purpose of this paper, instead of distinguishing them we will simply use the weight $w_k = w_{1,k} \, w_{0,k}$. 
$g$ is the inverse 5D phase-space volume which is constant along the particle trajectories. 
To obtain the corresponding grid distribution we must map the weights of all marker particles onto the grid. 
The velocity mapping in \xgc~is described in \citep[Sec.~III]{YoonPoP2014} and is made by moving a marker particle's weight between the particle position and the 2D uniform rectangular velocity-grid based on the inverse linear distance between the particle and the surrounding 4 grid points. This interpolation scheme 
has good enough particle number and parallel/perpendicular momentum conservation, but fails to preserve energy to a desired accuracy in some cases. 
In this section we explain why it is not straightforward numerically to extend this scheme, based on $\mathbb{P}_1$ Lagrange elements, to arbitrary order for conservation of higher-order moments. 
\subsection{1D uniform grid}\label{subsec:1DUniformGrid}
\begin{figure}
\begin{center}
	\includegraphics[width=0.75\textwidth]{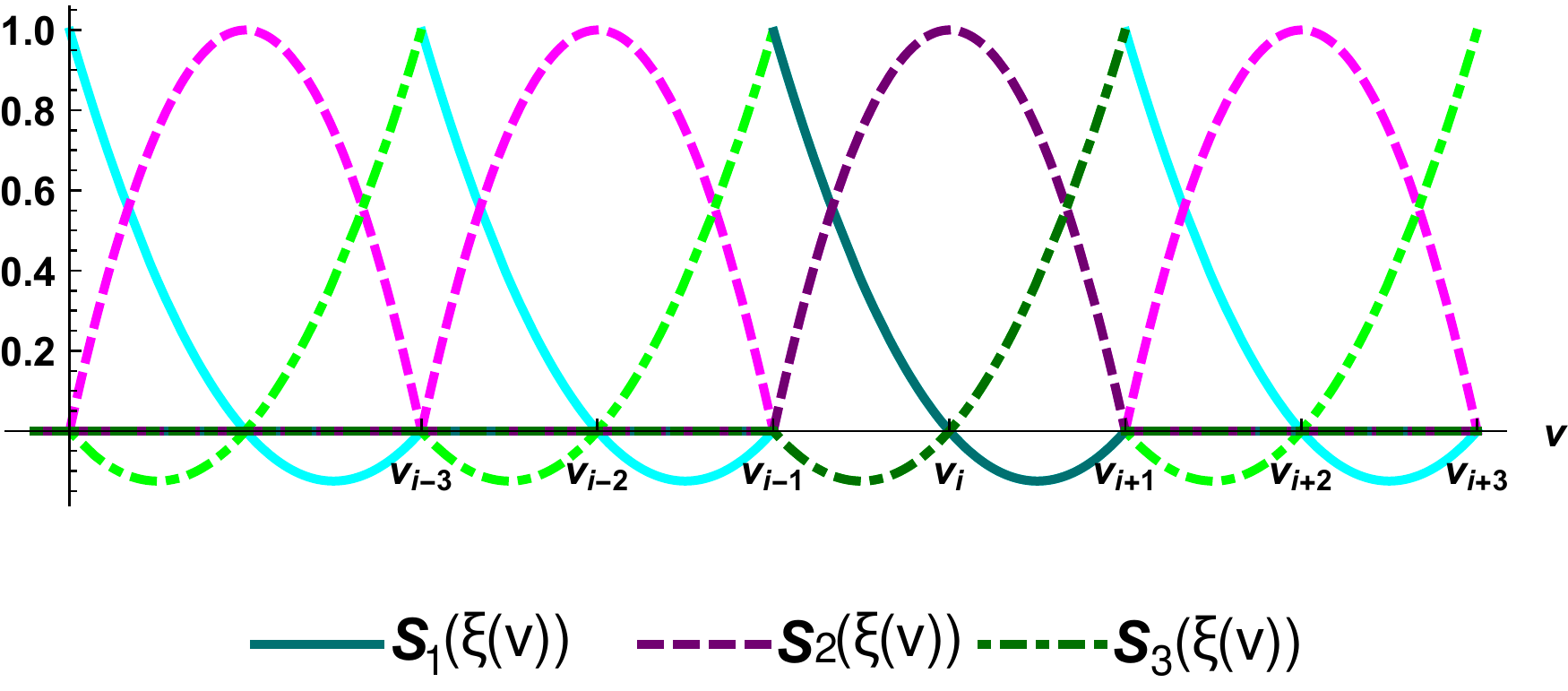}
\end{center}
\caption{Illustration of the shape functions $\displaystyle S_1 \left(\xi\left(v\right)\right), S_2 \left(\xi\left(v\right)\right), S_3 \left(\xi\left(v\right)\right)$ in 1D velocity space representing $\mathbb{P}_2$ Lagrange elements. 
}
\label{fig:QuadraticShapeFunctions1DPlot}
\end{figure}

It is enlightening to start by looking at shape functions on a 1D uniform grid. We wish to define a discrete representation of a continuous distribution function $f \left(v\right)$ on an interval $v \in \left[v_{\mathrm{min}}, v_{\mathrm{max}}\right]$. 
This can be achieved by a particle distribution of $N_p$ marker particles with weights $w_k$ and positions $v_k \in \left[v_{\mathrm{min}}, v_{\mathrm{max}}\right]$
\begin{equation}
f_{\mathrm{particle}}\left(v\right) = \sum_{k = 1}^{N_p} \frac{w_k}{V_k} \, \delta\left(v - v_k\right).
\label{eq:particleDistributionPIC1D}
\end{equation}
In Eq.~\eqref{eq:particleDistributionPIC1D} we have included the velocity-space volume $V_k$ represented by particle $k$, such that $f$ represents a distribution with number density $\displaystyle n = \int f \mathrm{d} v$. 
We can also represent $f \left(v\right)$ with a set of basis functions where the coefficients of the basis functions correspond to the function values on grid points. We write $\displaystyle f_{\mathrm{grid}, i} = f \left(v_i\right)$ using values at $N_g$ discrete grid points $v_1, v_2, \ldots, v_{N_g}$, evenly spaced with distance $\displaystyle \Delta v = v_{i} - v_{i - 1}$ and $v_1 = v_{\mathrm{min}}$, $v_{N_g} = v_{\mathrm{max}}$. 
In order to do $2^{\mathrm{nd}}$-order interpolation we assume an odd number of grid points, $N_g = 2 M + 1$ with $M = 0, 1, \ldots$, giving an even number of cells. 
To map the weight $w_k$ of a marker particle in Eq.~\eqref{eq:particleDistributionPIC1D} onto the 3 grid points $\displaystyle \left(v_{i-1}, v_i, v_{i+1}\right)$ of the $2^{\mathrm{nd}}$-order finite element in which the particle resides, we choose to use the $\mathbb{P}_2$ Lagrange shape functions represented by the polynomials 
\begin{equation}
\begin{array}{l}
	S_1 \left(\xi\right) = 1 - 3 \xi + 2 \xi^2  \\
	S_2 \left(\xi\right) = 4 \xi - 4 \xi^2 \\ 
	S_3 \left(\xi\right) = - \xi + 2 \xi^2 
\end{array}
\label{eq:2ndOrderLagrangeBasis}
\end{equation}
where $\displaystyle \xi \in [0, 1]$ and $0$ everywhere else. We define $\displaystyle \xi\left(v\right) = \frac{v - v_{i-1}}{v_{i+1} - v_{i-1}} = \frac{v - v_{i-1}}{2 \Delta v}$ for $v \in \left[v_{i-1}, v_{i+1}\right]$. 
The basis functions are illustrated in Fig.~\ref{fig:QuadraticShapeFunctions1DPlot}.
This choice conserves particle number density because $\displaystyle S_1 + S_2 + S_3 = 1; \forall \xi \in [0, 1]$. 
The particle number represented by the marker particle is
\begin{equation}
{\mathcal{N}}_{\mathrm{particle}, k} = w_k,
\label{eq:DensityParticle1D2ndOrderPolynomial}
\end{equation}
and the corresponding grid particle number, mapped from the particle, is 
\begin{multline}
{\mathcal{N}}_{\mathrm{grid}, i - 1} + {\mathcal{N}}_{\mathrm{grid}, i} + {\mathcal{N}}_{\mathrm{grid}, i + 1} = \\ = w_k S_1\left(\xi\left(v_k\right)\right) + w_k S_2\left(\xi\left(v_k\right)\right) + w_k S_3\left(\xi\left(v_k\right)\right) = w_k = {\mathcal{N}}_{\mathrm{particle}, k}.
\label{eq:DensityGrid1D2ndOrderPolynomial}
\end{multline}
To see if the $\mathbb{P}_2$ elements conserve momentum and energy we note the following identities:
\[
\begin{array}{l}
	S_3\left(\frac{v \, - \, v_{i - 1}}{2 \Delta v}\right) - S_1\left(\frac{v \, - \, v_{i - 1}}{2 \Delta v}\right) = \frac{v \, - \, v_i}{\Delta v}, \\
	S_1\left(\frac{v \, - \, v_{i - 1}}{2 \Delta v}\right) + S_3\left(\frac{v \, - \, v_{i - 1}}{2 \Delta v}\right) = \frac{\left(v \, - \, v_i\right)^2}{\left(\Delta v\right)^2}.
\end{array}
\]
The momentum of the marker particle is 
\begin{equation}
P_{\mathrm{particle}, k} = w_k v_k.
\label{eq:MomentumParticle1D2ndOrderPolynomial}
\end{equation}
The corresponding grid momentum is 
\begin{multline}
P_{\mathrm{grid}, i - 1} + P_{\mathrm{grid}, i} + P_{\mathrm{grid}, i + 1} = \\ = w_k \, S_1\left(\xi\left(v_k\right)\right) v_{i-1} + w_k \, S_2\left(\xi\left(v_k\right)\right) v_i + w_k \, S_3\left(\xi\left(v_k\right)\right) v_{i + 1} = \\
= w_k \left[\left(v_i - \Delta v\right) S_1 + v_i S_2 + \left(v_i + \Delta v\right) S_3 \right] = \\ = w_k \left[v_i \left(S_1 + S_2 +  S_3\right) + \Delta v \left(S_3 - S_1\right) \right] = \\
= w_k \left[v_i + \Delta v \frac{v_k - v_i}{\Delta v} \right] = w_k v_k = P_{\mathrm{particle}, k}.
\label{eq:MomentumGrid1D2ndOrderPolynomial}
\end{multline}
Analogously the energy of the marker particle is 
\begin{equation}
E_{\mathrm{particle}, k} = w_k v_k^2,
\label{eq:EnergyParticle1D2ndOrderPolynomial}
\end{equation}
and the corresponding grid energy 
\begin{multline}
E_{\mathrm{grid}, i - 1} + E_{\mathrm{grid}, i} + E_{\mathrm{grid}, i + 1} = \\ = w_k \, S_1\left(\xi\left(v_k\right)\right) v_{i-1}^2 + w_k \, S_2\left(\xi\left(v_k\right)\right) v_i^2 + w_k \, S_3\left(\xi\left(v_k\right)\right) v_{i + 1}^2 = \\
= w_k \left[\left(v_i - \Delta v\right)^2 S_1 + v_i^2 S_2 + \left(v_i + \Delta v\right)^2 S_3 \right] = \\ = w_k \left[v_i^2 \left(S_1 + S_2 +  S_3\right) + 2 v_i \Delta v \left(S_3 - S_1\right) + \left(\Delta v\right)^2 \left(S_1 + S_3\right) \right] = \\
= w_k \left[v_i^2 + 2 v_i \Delta v \frac{v_k - v_i}{\Delta v}  + \left(\Delta v\right)^2 \frac{\left(v_k - v_i\right)^2}{\left(\Delta v\right)^2} \right] = w_k v_k^2 = E_{\mathrm{particle}, k}.
\label{eq:EnergyPGrid1D2ndOrderPolynomial}
\end{multline}
We have thus shown that this $\mathrm{particles} \rightarrow \mathrm{grid}$ mapping preserves particle number, momentum and energy in 1D. 
This is in fact the only quadratic mapping of a single particle onto a uniform grid, with three non-zero shape functions in each finite element, 
that will simultaneously preserve density, momentum and energy, because $S_1 \left(\xi\right)$, $S_2 \left(\xi\right)$, $S_3 \left(\xi\right)$ in Eq.~\eqref{eq:2ndOrderLagrangeBasis} is the only solution to the system
\[
\begin{array}{ll}
	S_1\left(\xi\right) + S_2\left(\xi\right) + S_3\left(\xi\right) = 1 & \mathrm{(particle~conservation)}, \\
	0 \times S_1\left(\xi\right) + 0.5 \times S_2\left(\xi\right) + 1.0 \times S_3\left(\xi\right) = \xi & \mathrm{(momentum~conservation)}, \\
	0^2 \times S_1\left(\xi\right) + 0.5^2 \times S_2\left(\xi\right) + 1.0^2 \times S_3\left(\xi\right) = \xi^2 & \mathrm{(energy~conservation)}.
\end{array}
\]
\subsection{2D uniform grid}\label{subsec:2DUniformGrid}
\begin{figure}
\begin{center}
  \includegraphics[width=0.60\textwidth]{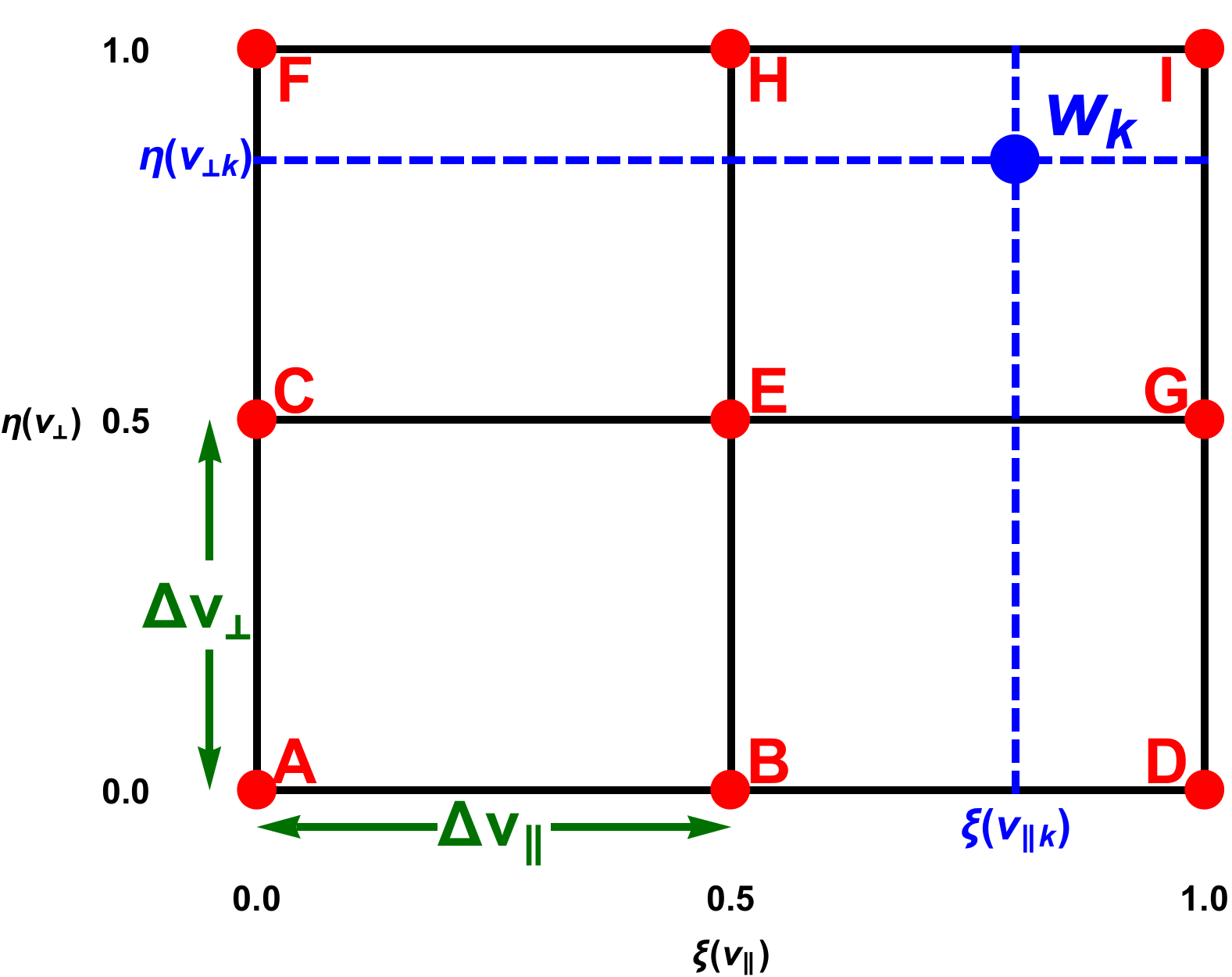}
\end{center}
\caption{Illustration of a finite element when using a quadratic basis in velocity space for a uniform rectangular grid in $\vpa$, $\vpe$, where we have labeled the 9~vertices A, B, ..., I.}
\label{fig:FiniteElement2D}
\end{figure}

In \xgc~the velocity space is a 2D uniform rectangular grid in $\vpa$, $\vpe$. The mapping onto the grid representation of the distribution function is described in \citep{YoonPoP2014} and is using the $\mathbb{P}_1$ Lagrange elements (1D shape functions $S_1\left(\xi\right) = 1 - \xi$; $S_2\left(\xi\right) = \xi$) which fail to preserve the energy of the marker particle representation. 
When extending the code to quadratic elements a finite element will encompass 9~vertices as illustrated in Fig.~\ref{fig:FiniteElement2D}, and we have to use an odd number of grid points in both dimensions. 
Analogously to the 1D basis functions in Eq.~\eqref{eq:2ndOrderLagrangeBasis}, 
we realize that we can map the weight $w_k$ of a marker particle onto the grid using the 9 2D-functions 
\begin{equation}
\begin{array}{l}
	K_{\mathrm{A}}\left(\xi, \eta\right) = S_1\left(\xi\right) \, S_1\left(\eta\right), \\
	K_{\mathrm{B}}\left(\xi, \eta\right) = S_2\left(\xi\right) \, S_1\left(\eta\right), \\
	K_{\mathrm{C}}\left(\xi, \eta\right) = S_1\left(\xi\right) \, S_2\left(\eta\right), \\
	K_{\mathrm{D}}\left(\xi, \eta\right) = S_3\left(\xi\right) \, S_1\left(\eta\right), \\
	K_{\mathrm{E}}\left(\xi, \eta\right) = S_2\left(\xi\right) \, S_2\left(\eta\right), \\
	K_{\mathrm{F}}\left(\xi, \eta\right) = S_1\left(\xi\right) \, S_3\left(\eta\right), \\
	K_{\mathrm{G}}\left(\xi, \eta\right) = S_3\left(\xi\right) \, S_2\left(\eta\right), \\
	K_{\mathrm{H}}\left(\xi, \eta\right) = S_2\left(\xi\right) \, S_3\left(\eta\right), \\
	K_{\mathrm{I}}\left(\xi, \eta\right) = S_3\left(\xi\right) \, S_3\left(\eta\right), 
\end{array}
\label{eq:2ndOrderPolynomialBasis2D}
\end{equation}
with $\displaystyle \xi\left(v_\|\right) = \frac{v_\| - v_{\|, i-1}}{v_{\|, i+1} - v_{\|, i-1}} \in \left[0, 1\right]$ and $\displaystyle \eta\left(v_\perp\right) = \frac{v_\perp - v_{\perp, j-1}}{v_{\perp, j+1} - v_{\perp, j-1}} \in \left[0, 1\right]$ in the 2D velocity cell where the particle resides, and where we have labeled the grid vertices as in Fig.~\ref{fig:FiniteElement2D}. 

To extend the bilinear mapping in \citep{YoonPoP2014} to a quadratic mapping we define the coefficients
\begin{multline}
\mathcal{P}_{k \rightarrow \mathrm{A}} = K_{\mathrm{A}}\left(\xi\left(v_{\| k}\right), \eta\left(v_{\perp k}\right)\right), \;
\ldots,
\mathcal{P}_{k \rightarrow \beta} = K_{\beta}\left(\xi\left(v_{\| k}\right), \eta\left(v_{\perp k}\right)\right), \\
\ldots,
\mathcal{P}_{k \rightarrow \mathrm{I}} = K_{\mathrm{I}}\left(\xi\left(v_{\| k}\right), \eta\left(v_{\perp k}\right)\right)
\label{eq:2ndOrderPolynomial2Dmapping}
\end{multline}
describing the fraction of weight distributed from particle $k$ onto grid vertex $\beta$. 
For the $N_p$ marker particles residing within a 2D cell the velocity space representation of the distribution can be expressed as 
\begin{equation}
f_{\mathrm{particle}}\left(v_\|, v_\perp\right) = \sum_{k = 1}^{N_p} \frac{w_k}{V_k} \, \delta\left(v_\| - v_{\| k}, v_\perp - v_{\perp k}\right).
\label{eq:particleDistribution2D}
\end{equation}
We can then use the mapping in Eq.~\eqref{eq:2ndOrderPolynomial2Dmapping} to obtain the corresponding grid distribution of vertex $\beta$ as
\begin{equation}
f_{\mathrm{grid}, \beta} = \frac{1}{V_{\beta}} \sum_{k = 1}^{N_p} w_k \mathcal{P}_{k \rightarrow \beta},
\label{eq:meshDistribution2D}
\end{equation}
where 
$\displaystyle V_{\beta} = 2 \upi v_{\perp \beta} \Delta v_{\perp} \Delta v_{\|}$ is the velocity-space volume of the vertex. 
It is straightforward 
to verify that this mapping conserves particle number
\[
\mathcal{P}_{k \rightarrow \mathrm{A}} + \mathcal{P}_{k \rightarrow \mathrm{B}} + \ldots + \mathcal{P}_{k \rightarrow \mathrm{I}} = 1, 
\]
parallel momentum
\[
v_{\| \mathrm{A}} \mathcal{P}_{k \rightarrow \mathrm{A}} + v_{\| \mathrm{B}} \mathcal{P}_{k \rightarrow \mathrm{B}} + \ldots + v_{\| \mathrm{I}} \mathcal{P}_{k \rightarrow \mathrm{I}} = v_{\| k},
\]
perpendicular momentum
\[
v_{\perp \mathrm{A}} \mathcal{P}_{k \rightarrow \mathrm{A}} + v_{\perp \mathrm{B}} \mathcal{P}_{k \rightarrow \mathrm{B}} + \ldots + v_{\perp \mathrm{I}} \mathcal{P}_{k \rightarrow \mathrm{I}} = v_{\perp k},
\]
and energy
\[
\left(v_{\| \mathrm{A}}^2 + v_{\perp \mathrm{A}}^2\right) \mathcal{P}_{k \rightarrow \mathrm{A}} + \left(v_{\| \mathrm{B}}^2 + v_{\perp \mathrm{B}}^2\right) \mathcal{P}_{k \rightarrow \mathrm{B}} + \ldots + \left(v_{\| \mathrm{I}}^2 + v_{\perp \mathrm{I}}^2\right) \mathcal{P}_{k \rightarrow \mathrm{I}} = v_{\| k}^2 + v_{\perp k}^2,
\]
but we omit the details here. 

After the Coulomb collision operation, and other source operations, we obtain an updated grid distribution $\displaystyle f_{\mathrm{grid}, \beta}^{\mathrm{new}}$. It is possible to do the inverse mapping back from vertex $\beta$ to particle $\alpha$ using the coefficient 
\begin{equation}
\mathcal{P}_{\beta \rightarrow \alpha}^{(-1)} = \frac{\mathcal{P}_{\alpha \rightarrow \beta}}{\sum_{k = 1}^{N_p} \mathcal{P}_{k \rightarrow \beta}},
\label{eq:2ndOrderPolynomial2DinverseMapping}
\end{equation}
and update the new weight of the particle as 
\begin{equation}
w_\alpha^{\mathrm{new}} = \sum_{\beta} f_{\mathrm{grid}, \beta}^{\mathrm{new}} V_{\beta} \mathcal{P}_{\beta \rightarrow \alpha}^{(-1)}.
\label{eq:2ndOrderPolynomial2DNewWeight}
\end{equation}
This inverse mapping is a direct extension of the bilinear scheme in \xgc. Note that Eq.~\eqref{eq:2ndOrderPolynomial2DinverseMapping} is slightly different from the corresponding equation in \citep{YoonPoP2014} which also contains the marker particle weights. This modification was introduced when moving from the ``full-f'' to the ``total-$\delta f$'' as the default version of \xgc, since the weighting is more suitable in a ``full-f'' scheme.

Although the presented quadratic mapping theoretically will work, it has two major drawbacks. Firstly, the inverse mapping given by Eqs.~\eqref{eq:2ndOrderPolynomial2DinverseMapping} and \eqref{eq:2ndOrderPolynomial2DNewWeight} will not result in exact momentum and energy conservation when moving from the grid representation to the marker particle representation. Secondly, and more severely, 
in contrast to the bilinear scheme, the mapping coefficients in Eq.~\eqref{eq:2ndOrderPolynomial2Dmapping} can be negative. As a consequence, when doing the inverse mapping the denominator in Eq.~\eqref{eq:2ndOrderPolynomial2DinverseMapping} can become arbitrarily close to 0 resulting in numerical issues. 
In the next section we will introduce a way to overcome these issues by using a pseudo-inverse. 

The reader may question why negative mapping coefficients are allowed at all since this can lead to a negative grid distribution according to Eq.~\eqref{eq:meshDistribution2D}. However, a negative grid distribution is not a problem as long as the total distribution function (including the analytic part) in Eq.~\eqref{eq:XGCtotalDistribution} stays positive. In fact, the $w_{1,k}$ weights, described after Eq.~\eqref{eq:deltafDistribution}, are free to have either sign in \xgc~also with a bilinear velocity-space mapping. The non-Maxwellian part of the total distribution function can be as large as the Maxwellian background and if $f_s$ in Eq.~\eqref{eq:XGCtotalDistribution} becomes negative, \xgc~implements a technique to handle this to prevent unphysical values of the density, flow and temperature. This situation typically only occurs if the particle noise is large in regions with low background $f_0$, which can happen e.g. when particles enter the SOL from the pedestal. 
%
\section{Novel velocity mapping technique}\label{sec:NovelTechnique}
In this section we will describe a different method to map marker particle weights back-and-forth between particle positions and the velocity grid which avoids the problem of singularities described in the previous section, and enables exact conservation of moments up to the order of the finite element space. 
Assume a particle distribution of $N_p$ particles
\begin{equation}
\overbar{f}_{\mathrm{particle}}\left(\boldsymbol{v}\right) = \sum_{k = 1}^{N_p} w_k \, \delta\left(\boldsymbol{v} - {\boldsymbol{v}_{k}}\right),
\label{eq:particleDistributionPIC2D}
\end{equation}
a finite element space $\mathcal{F}$ and a function $u \in \mathcal{F}$. 
Since we will effectively be mapping particle weights rather than particle densities, we choose to omit the velocity-space volume present in Eqs.~\eqref{eq:particleDistributionPIC1D} and \eqref{eq:particleDistribution2D} in the definition in Eq.~\eqref{eq:particleDistributionPIC2D}. Hence the overbar notation to distinguish it from a regular distribution function. 
We would like to have a unique $\displaystyle u = \overbar{f}_{\mathrm{particle}}\left(\boldsymbol{v}\right);~\forall\,\boldsymbol{v}$, but this is not possible unless $\overbar{f}_{particle} \in \mathcal{F}$. Instead we write
\begin{equation}
u = \sum_{j} c_j \phi_j,
\label{eq:uFunction}
\end{equation}
where the functions $\phi_j \left(\boldsymbol{v}\right)$ form a complete basis of $\mathcal{F}$, and look for a solution that fulfills weak equivalence
\begin{equation}
\int \mathrm{d}\boldsymbol{v} \phi_i \, \overbar{f}_{\mathrm{particle}} = \int \mathrm{d}\boldsymbol{v} \phi_i \, u
\label{eq:uWeakEquivalence}
\end{equation}
for all $\phi_i \in \mathcal{F}$ (i.e. we find the projection of $\overbar{f}_{\mathrm{particle}}$ onto the function space $\mathcal{F}$). 
Note that the left-hand-side of Eq.~\eqref{eq:uWeakEquivalence} is simply obtained by evaluating the particle positions in the basis functions. 
We define a finite element mass matrix
\begin{equation}
\mathsfbi{M}_{ij} = \int \mathrm{d}\boldsymbol{v} \phi_i \phi_j,
\label{eq:MassMatrix}
\end{equation}
and a matrix set by the particle positions
\begin{equation}
\mathsfbi{V}_{jk} = \phi_j \left(\boldsymbol{v}_{k}\right).
\label{eq:ParticlePositionMatrix}
\end{equation}
We can then write weak equivalence as
\begin{equation}
\mathsfbi{M} \boldsymbol{c} = \mathsfbi{V} \boldsymbol{w},
\label{eq:WeakEquivalenceMatrixForm}
\end{equation}
where we have defined the vector $\boldsymbol{w}$ by the particle weights in Eq.~\eqref{eq:particleDistributionPIC2D} and the vector $\boldsymbol{c}$ by the coefficients in Eq.~\eqref{eq:uFunction} \citep{HirvijokiPoP2017}. 
$\mathsfbi{M}$ is a sparse invertible square matrix of size equal to the number of basis functions, which in practice is equal to the number of grid points on the velocity grid, $\displaystyle N_g \times N_g$. 
The sparsity follows from the fact that most pairs of basis functions have no overlapping support. 
Similarly $\mathsfbi{V}$ is also a sparse matrix because a particle position evaluated in a basis function will only be non-zero for the basis functions of the element in which the particle resides. $\mathsfbi{V}$ is rectangular of size $\displaystyle N_g \times N_p$ generally with the same number of non-zero matrix entries per column as there are grid vertices per element (i.e. 4 for linear elements or 9 for quadratic elements) and each column sums to 1. 
In the typical case of more particles than grid points, $N_p > N_g$, $\mathsfbi{V}$ is a short fat matrix (in our implementation described here this is practically always the case due to the particle resampling described later), but it could in principle also be a square or tall skinny matrix if the particle count is very low. 
To map the weights from the marker particle representation of the distribution function to the grid representation we simply do
\begin{equation}
\overbar{f}_{\mathrm{grid}} = \mathsfbi{V} \boldsymbol{w},
\label{eq:gridDistributionPIC}
\end{equation} 
and the coefficients of the basis functions become
\begin{equation}
\boldsymbol{c} = \mathsfbi{M}^{-1} \mathsfbi{V} \boldsymbol{w}.
\label{eq:gridCoefficients}
\end{equation} 
After applying grid operations we note from Eq.~\eqref{eq:WeakEquivalenceMatrixForm} that it would be possible to map back to the particle representation using $\displaystyle \mathsfbi{V}^{\mathrm{T}} \mathsfbi{M} \boldsymbol{c}_{\mathrm{new}}$, where the superscript ``$\mathrm{T}$'' denotes the transpose and the subscript ``new'' refers to after the grid operations, but this is not preserving the moments. 
Instead we would like to compute the new particle weights $\boldsymbol{w}_{\mathrm{new}}$ using an inverse of $\mathsfbi{V}$, but this inverse is not well defined. We have to resort to some kind of pseudo-inverse, $\mathsfbi{V}^{+}$. 
%
In the (unusual) case of fewer particles than grid points, $N_p < N_g$, we can use 
\begin{equation}
\mathsfbi{V}^{+} =  \left(\mathsfbi{V}^{\mathrm{T}} \mathsfbi{V}\right)^{-1} \mathsfbi{V}^{\mathrm{T}},
\label{eq:PseudoInverseLeft}
\end{equation}
which is a left inverse since $\displaystyle \mathsfbi{V}^{+} \mathsfbi{V} = \mathsfbi{I}$, the identity matrix.
If we were to map back-and-forth using the left inverse with an identity operation on the grid, e.g. no operation, we would end up with the new weights $\displaystyle \boldsymbol{w}_{\mathrm{new}} = \left(\mathsfbi{V}^{\mathrm{T}} \mathsfbi{V}\right)^{-1} \mathsfbi{V}^{\mathrm{T}} \mathsfbi{V} \boldsymbol{w} = \boldsymbol{w}$, i.e. we would recover the original weights without loosing any information.
However, for the typical case of more particles than grid points, $N_p > N_g$, we have to use the right inverse, $\displaystyle \mathsfbi{V} \mathsfbi{V}^{+} = \mathsfbi{I}$, given by the Moore-Penrose inverse \citep{Penrose1955} 
\begin{equation}
\mathsfbi{V}^{+} =  \mathsfbi{V}^{\mathrm{T}} \left(\mathsfbi{V} \mathsfbi{V}^{\mathrm{T}}\right)^{-1},
\label{eq:PseudoInverseRight}
\end{equation}
and update the new particle weights as
\begin{equation}
\boldsymbol{w}_{\mathrm{new}} = \mathsfbi{V}^{\mathrm{T}} \left(\mathsfbi{V} \mathsfbi{V}^{\mathrm{T}}\right)^{-1} \mathsfbi{M} \boldsymbol{c}_{\mathrm{new}}.
\label{eq:PseudoInverse}
\end{equation}
This transformation preserves the moments up to the order of the finite elements \citep{HirvijokiArxiv2018}. 
If we again were to map back-and-forth with an identity operation on the grid we would end up with the new weights $\displaystyle \boldsymbol{w}_{\mathrm{new}} = \mathsfbi{V}^{\mathrm{T}} \left(\mathsfbi{V} \mathsfbi{V}^{\mathrm{T}}\right)^{-1} \mathsfbi{V} \boldsymbol{w}$. This implies that the particle weights have changed, i.e. although the moments are conserved we will unavoidably loose information. 
The fact that we will generally loose information when $N_p > N_g$, unless there is a specific structure in the particle data that we can exploit, follows from the impossibility to always recover a larger set of data in the particles from a smaller set of data in the grid points (i.e. we cannot use the left inverse when $N_p > N_g$ except if the operand happens to be in the span of $\mathsfbi{V}^{\mathrm{T}}$).
If we apply the right pseudo-inverse mapping back-and-forth twice with an identity grid operation we obtain $\displaystyle \boldsymbol{w}_{\mathrm{new}} = \mathsfbi{V}^{\mathrm{T}} \left(\mathsfbi{V} \mathsfbi{V}^{\mathrm{T}}\right)^{-1} \mathsfbi{V} \mathsfbi{V}^{\mathrm{T}} \left(\mathsfbi{V} \mathsfbi{V}^{\mathrm{T}}\right)^{-1} \mathsfbi{V} \boldsymbol{w} = \mathsfbi{V}^{\mathrm{T}} \left(\mathsfbi{V} \mathsfbi{V}^{\mathrm{T}}\right)^{-1} \mathsfbi{V} \boldsymbol{w}$, i.e. the same result as when mapping only once. The process is idempotent since no more information is lost when applying the operation several times. 

A numerical implementation of the described mapping technique with the pseudo-inverse has been included in the \petsc~library, version 3.14 \citep{petsc2020}. We use \petsc~to implement a new velocity interpolation in \xgc~with the linear $1^{\mathrm{st}}$-order $\mathbb{P}_1$ Lagrange elements or the quadratic $2^{\mathrm{nd}}$-order $\mathbb{P}_2$ elements as basis functions. It would be a straightforward extension to implement even higher-order elements, but our target is to achieve energy conservation for which $2^{\mathrm{nd}}$-order elements suffice -- as is the case for the \xgc~simulations. 
The `$\mathrm{particles} \rightarrow \mathrm{grid}$' mapping merely consists of constructing the $\mathsfbi{M}$, $\mathsfbi{V}$ matrices, the weights vector $\boldsymbol{w}$, and performing matrix multiplications. 
With $\mathbb{P}_1$ elements this operation gives the same result as the former \xgc~velocity interpolation, whereas $\mathbb{P}_2$ elements result in a mapping equivalent to Eq.~\eqref{eq:meshDistribution2D} (taking the velocity-space volumes into account). 
The inverse `$\mathrm{grid} \rightarrow \mathrm{particles}$' mapping is more critical because it involves constructing the pseudo-inverse using an iterative Krylov solver in \petsc. This operation can fail 
if $\displaystyle \left(\mathsfbi{V} \mathsfbi{V}^{\mathrm{T}}\right)^{-1}$ in Eq.~\eqref{eq:PseudoInverse} is not well posed, which can happen (but not necessarily) when there are basis functions without any particles.
To ensure that the pseudo-inverse calculation will always work we have to ensure that all cells on the velocity grid, in all real-space mesh vertices, are populated by particles. In \xgc~simulations, with a realistic number of marker particles and a typical 5D grid size, we expect that a fraction $\sim1\%$ of all velocity cells are not populated during a time step, and particularly cells at high kinetic energy. 
Appropriately, \xgc~is equipped with a resampling technique which is used to redistribute particles in velocity space while preserving a desired number of moments \citep{FaghihiJCP2020}. 
\citep[More details on the \xgc~resampling in][]{DominskiPoP2020}.
We apply the resampling just before the `$\mathrm{particles} \rightarrow \mathrm{grid}$' mapping to ensure that the pseudo-inverse can always be calculated. Since the resampling will fill all empty velocity cells with particles we will practically never end up in a situation where $N_p < N_g$ (except in extremely special cases like when there is only one single particle in each velocity cell on the grid) and the left pseudo-inverse in Eq.~\eqref{eq:PseudoInverseLeft} is consequently not used.
%
\section{Application to a neoclassical XGC simulation}\label{sec:Application}
\begin{figure}
\begin{center}
  \includegraphics[width=0.9\textwidth]{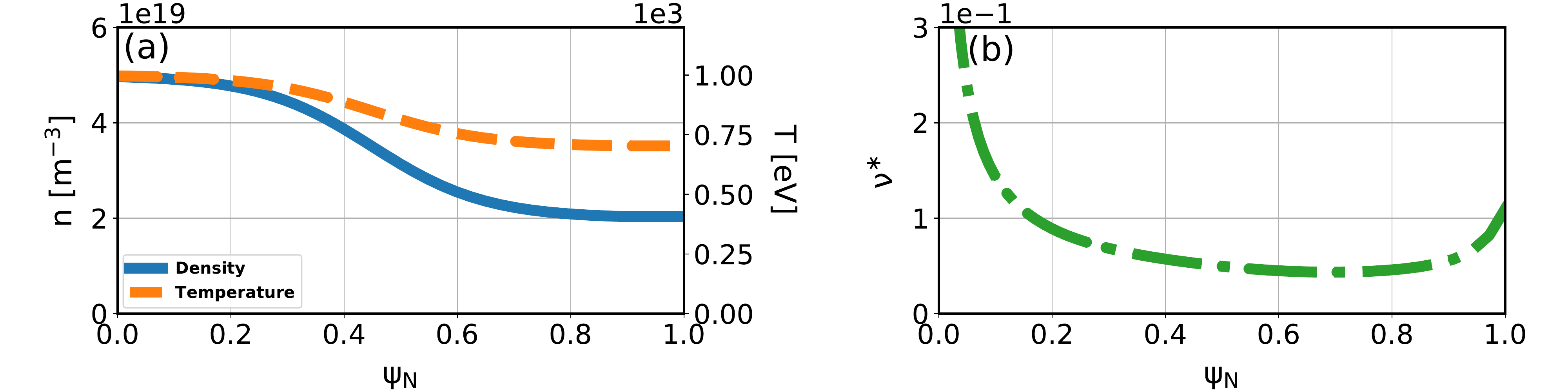}
\end{center}
\caption{
Input plasma profiles of ion density [\textcolor{NewBlue}{\newfull}], temperature~[\textcolor{NewOrange}{\longdashed}]~(a) and collisionality [\textcolor{NewGreen}{\dashdot}] (b).}
\label{fig:PlasmaProfiles}
\end{figure}

Here we apply the new velocity mapping described in Sec.~\ref{sec:NovelTechnique} to a circular cross-sectional tokamak neoclassical test case for \xgc, the same case as in \citep{HagerPoP2019} but for simplicity we use adiabatic instead of kinetic electrons, and demonstrate the improved conservation properties of the interpolation. Note that no significant influence of the different interpolation methods on the physical observables are expected in this test case due to its mild density and temperature gradients. Our objective is rather to demonstrate that the new implementation reduces the interpolation error to a negligible level. Therefore, we will focus on the collision- and interpolation errors. 
Even if the interpolation errors are too small to influence the physical simulation results here, it is always desirable to have interpolation errors that are smaller than the collision errors. In realistic edge cases with steep pedestal profiles the interpolation errors are typically larger and can cause numerical issues if they are too large. 
In the comparison we will refer to the former \xgc~velocity mapping as the ``bilinear'' implementation, whereas we call the new mapping $1^{\mathrm{st}}$- and $2^{\mathrm{nd}}$-order ``pseudo-inverse'' respectively. 
We will omit the species index in the presented results because all quantities are for the main deuterium ions with mass $m$. The input plasma profiles of density $n$, temperature $T$ and collisionality at the beginning of the simulations are illustrated in Fig.~\ref{fig:PlasmaProfiles} as a function of normalized poloidal flux $\psi_N$. The collisionality is defined as $\displaystyle {\nu}^{\ast} = q R \nu / \left({\epsilon}^{3/2} v_{T}\right)$ with safety factor $q$, major radius $R$, inverse aspect ratio $\epsilon$, thermal velocity $v_{T} = \sqrt{2 T / m}$ and collision frequency $\nu$ \citep{helander02collisional}. The profiles are free to evolve during a simulation, but because the simulated time is much shorter than the neoclassical core transport timescale they remain practically unchanged. The 3D configuration space mesh consists of 6368 nodes, the velocity grid is $45 \times 45$ grid points, and the simulation domain is set to $\psi_N < 0.9$. 
In neoclassical tokamak \xgc~simulations a single poloidal plane is used because of axisymmetry.
The total number of particles is $\sim 30$ million throughout the simulations but varies $\pm 5$\% because of the resampling. We run 400 time steps of length $\Delta t = 0.28  \mathrm{\mu s}$ and study the errors.
%
%
%

We define the collisional change in density, parallel momentum, and energy during a time step of the simulation as
\begin{equation}
\begin{array}{l}
	\Delta n = \Delta t \int C_{ii} \mathrm{d}^3 v,\\
	\Delta P_{\|} = \Delta t \int m \vpa C_{ii} \mathrm{d}^3 v,\\
	\Delta E = \Delta t \int \frac{1}{2} m v^2 C_{ii} \mathrm{d}^3 v,
\end{array}
\label{eq:collisionalChange}
\end{equation}
where $\displaystyle C_{ii} = \left(\frac{\mathrm{d}f}{\mathrm{d}t}\right)_{\mathrm{coll}}$ is the ion-ion collision operator acting on the distribution function. 
The Fokker-Planck-Landau collision operator implemented in \xgc~conserves particle density ($\displaystyle \Delta n = 0$), parallel momentum ($\displaystyle \Delta P_{\|} = 0$) and kinetic energy ($\displaystyle \Delta E = 0$) up to a user-defined accuracy for a single ion species simulation (in a multi-species simulation all inter-species collisions must also be considered and the total momentum and energy are conserved instead). 
We therefore let the numerical error of a quantity be the absolute value of the corresponding change in the quantity (however, in color map plots we leave the sign). 
The collision operation in the code is implemented with an implicit time discretization that is solved iteratively until the desired conservation error tolerances are met. This implies that the collision errors can be made arbitrarily small down to machine precision by choosing stricter tolerances (and potentially a smaller time step). However, since the collision operation can be a substantial fraction of the total computing time the tolerances are set such that the errors are small enough to limit the number of iterations while keeping the accumulated conservation error at an acceptable level over the required number of time steps.  
In contrast to the velocity interpolation, the collision operator does not conserve perpendicular momentum. 
While the collision errors are calculated by looking at the changes on the velocity grid, the interpolation errors have to be calculated using the marker particle representation. The interpolation errors are therefore computed using the total changes of mass, momentum, and energy in the full process `$\mathrm{particles} \rightarrow \mathrm{grid}$' mapping, collisions, `$\mathrm{grid} \rightarrow \mathrm{particles}$' mapping, and then subtracting the collision errors.

\begin{figure}
\begin{center}
  \includegraphics[width=0.7\textwidth]{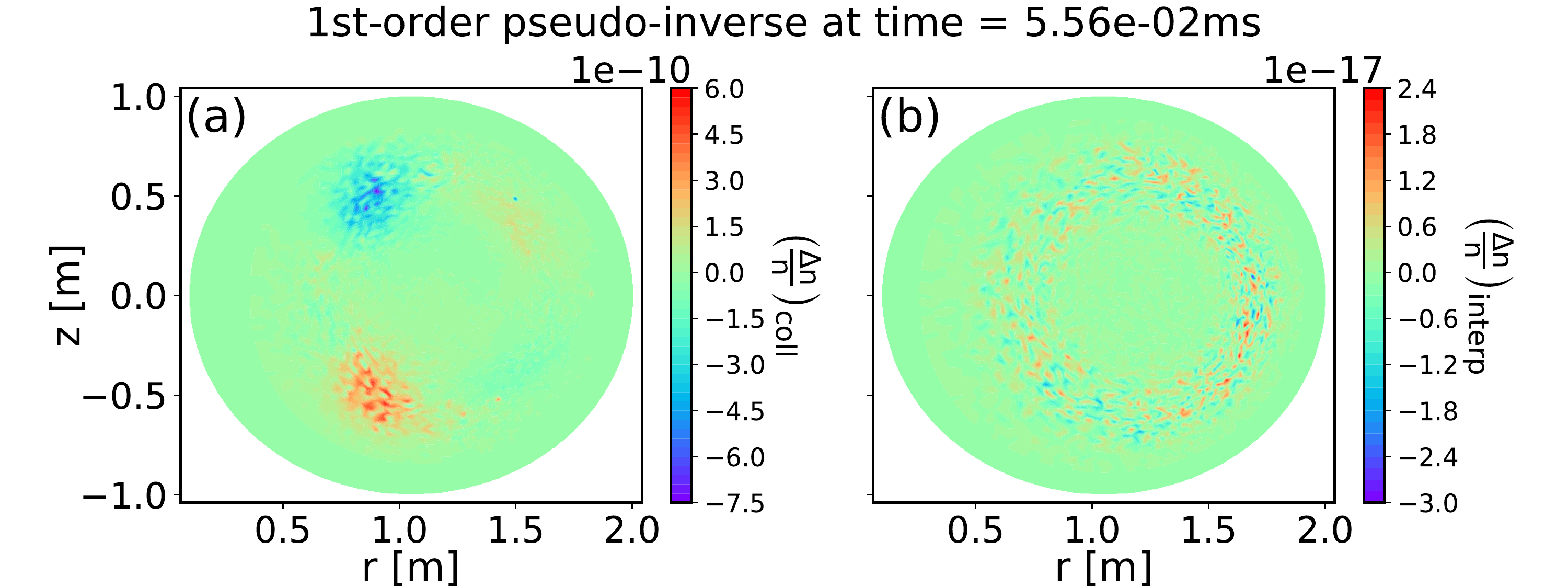}\\ \vspace*{0.2cm}
  \includegraphics[width=0.7\textwidth]{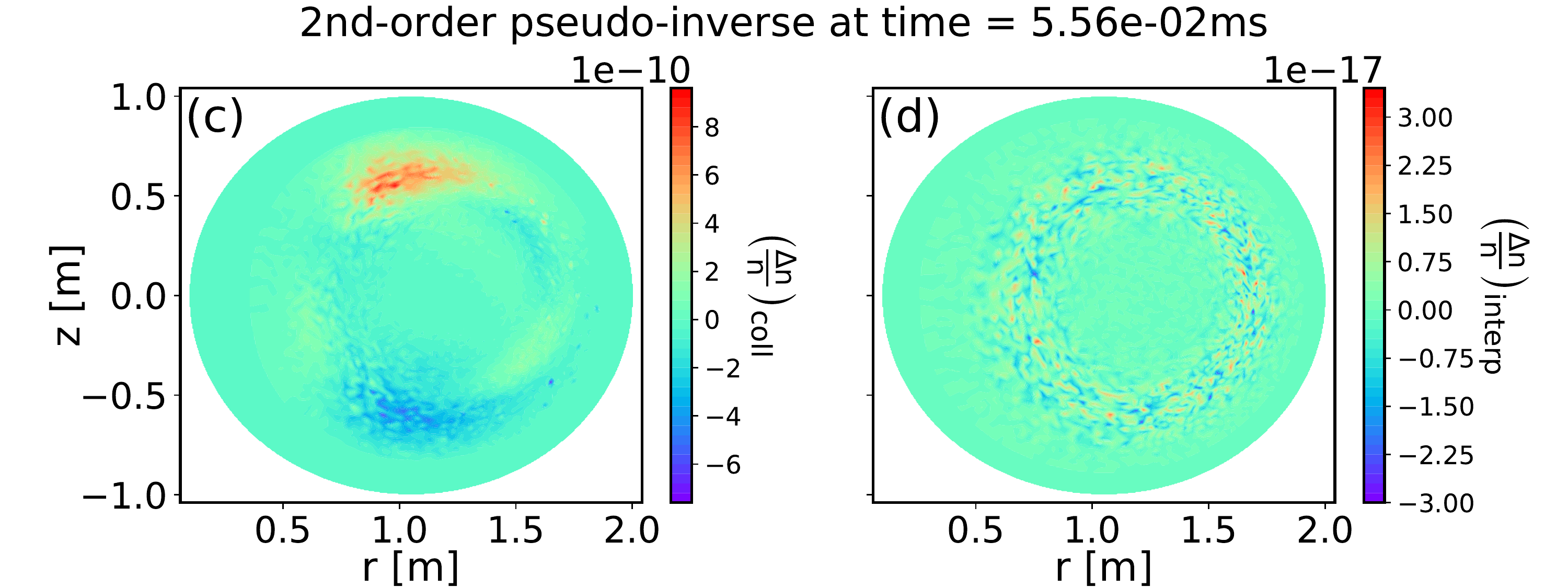}\\ \vspace*{0.2cm}
  \includegraphics[width=0.7\textwidth]{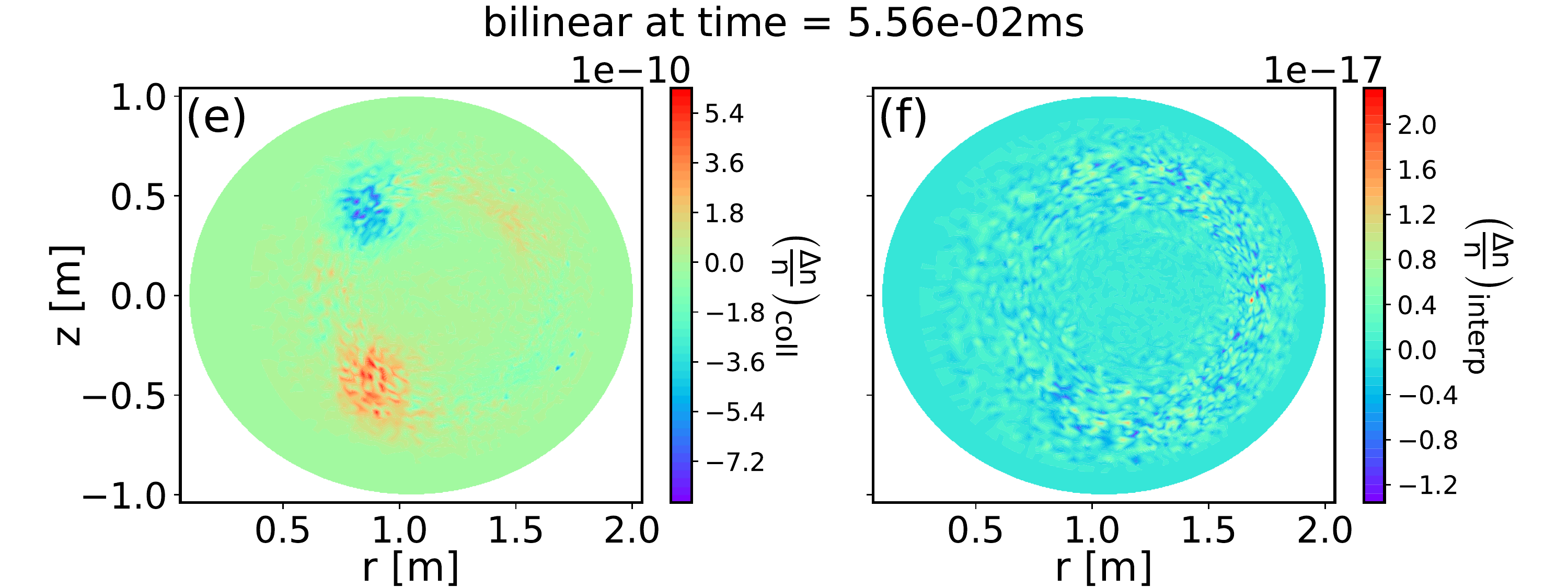}
\end{center}
\caption{
Color maps of the change in relative ion density $\displaystyle \Delta n / n$ (relative density error) in all configuration space vertices of the poloidal cross-section in time step 200 of the simulation. The color maps show the change due to the collisions (a, c, e) and the change due to the interpolation (b, d, f), comparing the three different velocity mapping methods: $1^{\mathrm{st}}$-order ``pseudo-inverse'' (a, b), $2^{\mathrm{nd}}$-order ``pseudo-inverse'' (c, d) and ``bilinear'' (e, f).}
\label{fig:densityError}
\end{figure}

\begin{figure}
\begin{center}
  \includegraphics[width=0.7\textwidth]{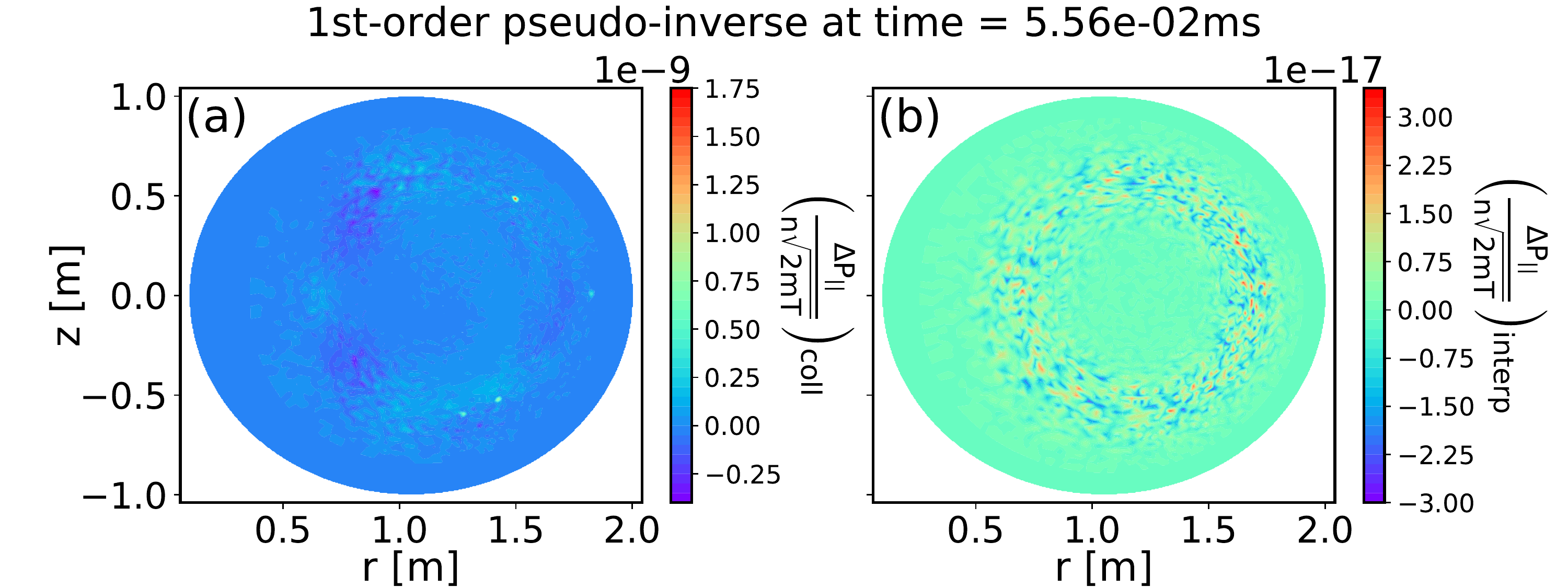}\\ \vspace*{0.2cm}
  \includegraphics[width=0.7\textwidth]{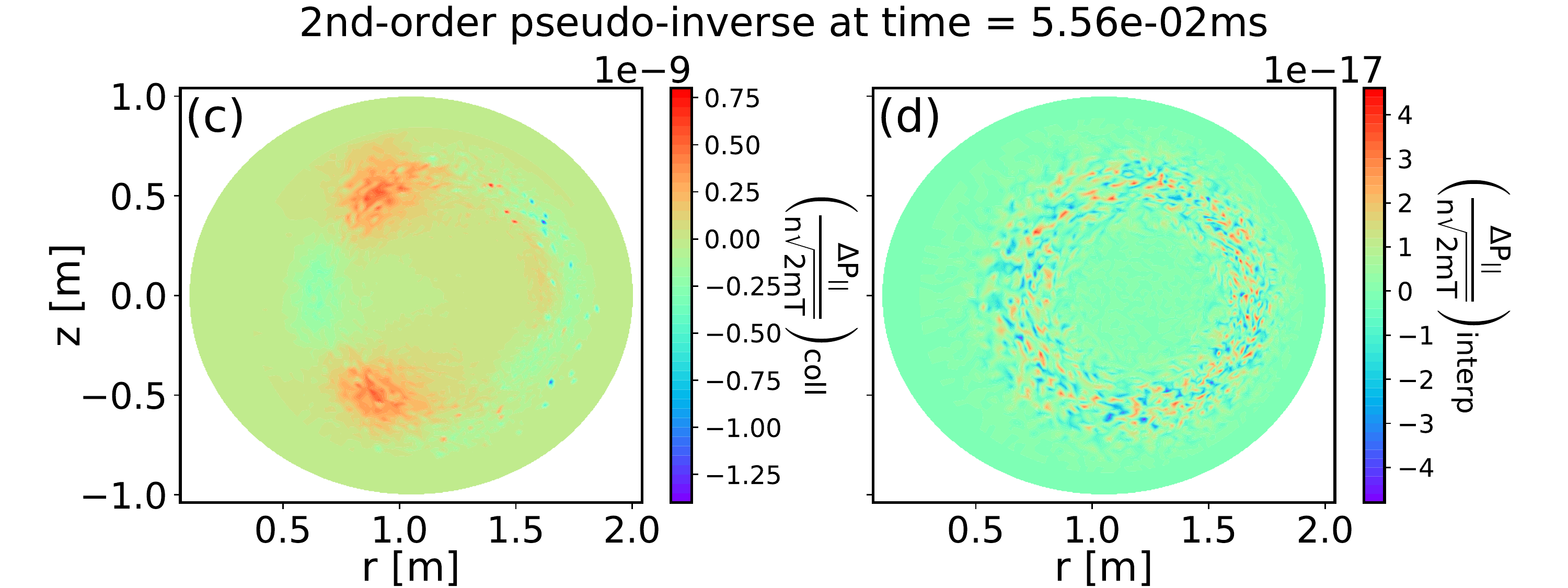}\\ \vspace*{0.2cm}
  \includegraphics[width=0.7\textwidth]{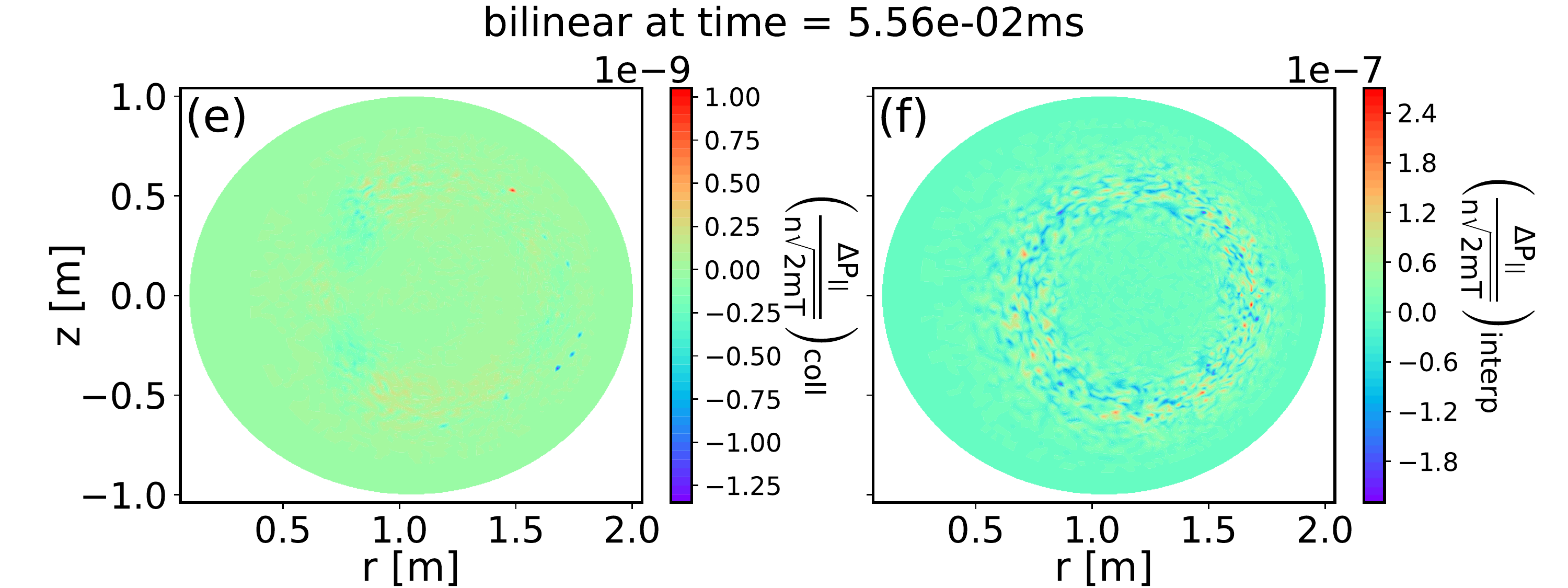}
\end{center}
\caption{
Color maps of the change in linear parallel momentum normalized to momentum at thermal speed $\displaystyle \frac{\Delta P_{\|}}{n \sqrt{2 m T}}$ (relative parallel momentum error) for the ions in all configuration space vertices of the poloidal cross-section in time step 200 of the simulation. The color maps show the change due to the collisions (a, c, e) and the change due to the interpolation (b, d, f), comparing the three different velocity mapping methods: $1^{\mathrm{st}}$-order ``pseudo-inverse'' (a, b), $2^{\mathrm{nd}}$-order ``pseudo-inverse'' (c, d) and ``bilinear'' (e, f).}
\label{fig:momentumError}
\end{figure}

\begin{figure}
\begin{center}
  \includegraphics[width=0.7\textwidth]{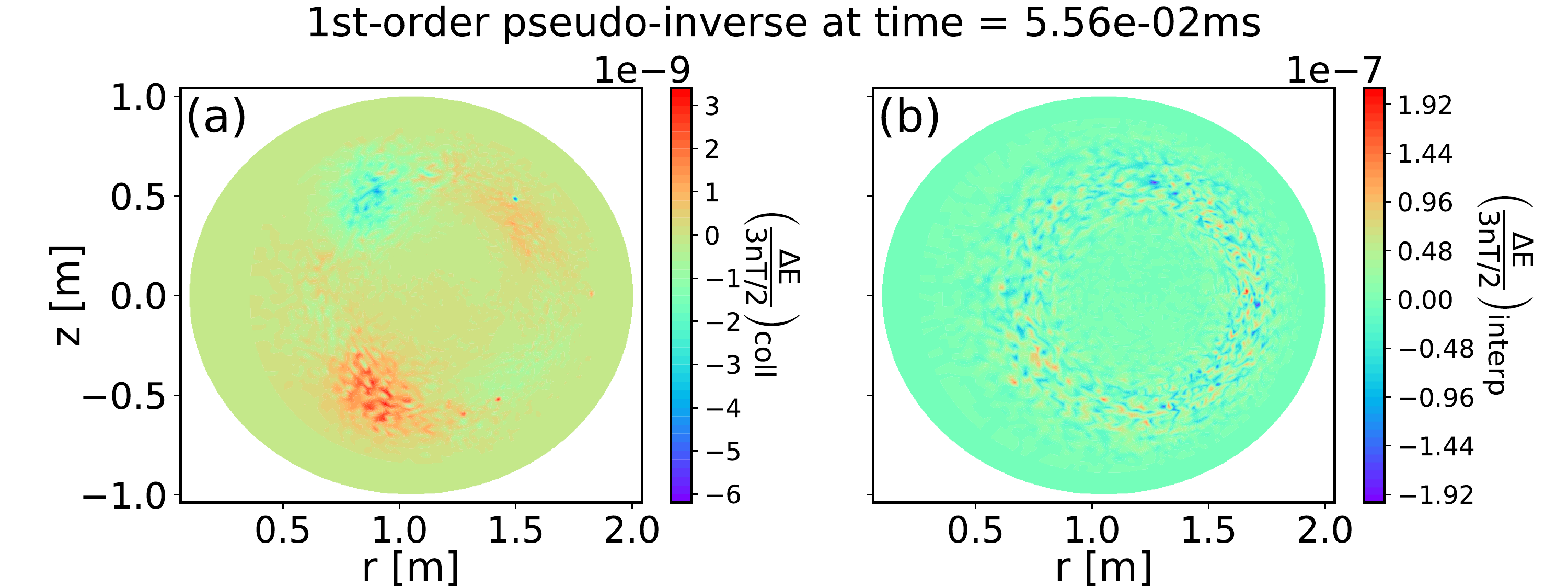}\\ \vspace*{0.2cm}
  \includegraphics[width=0.7\textwidth]{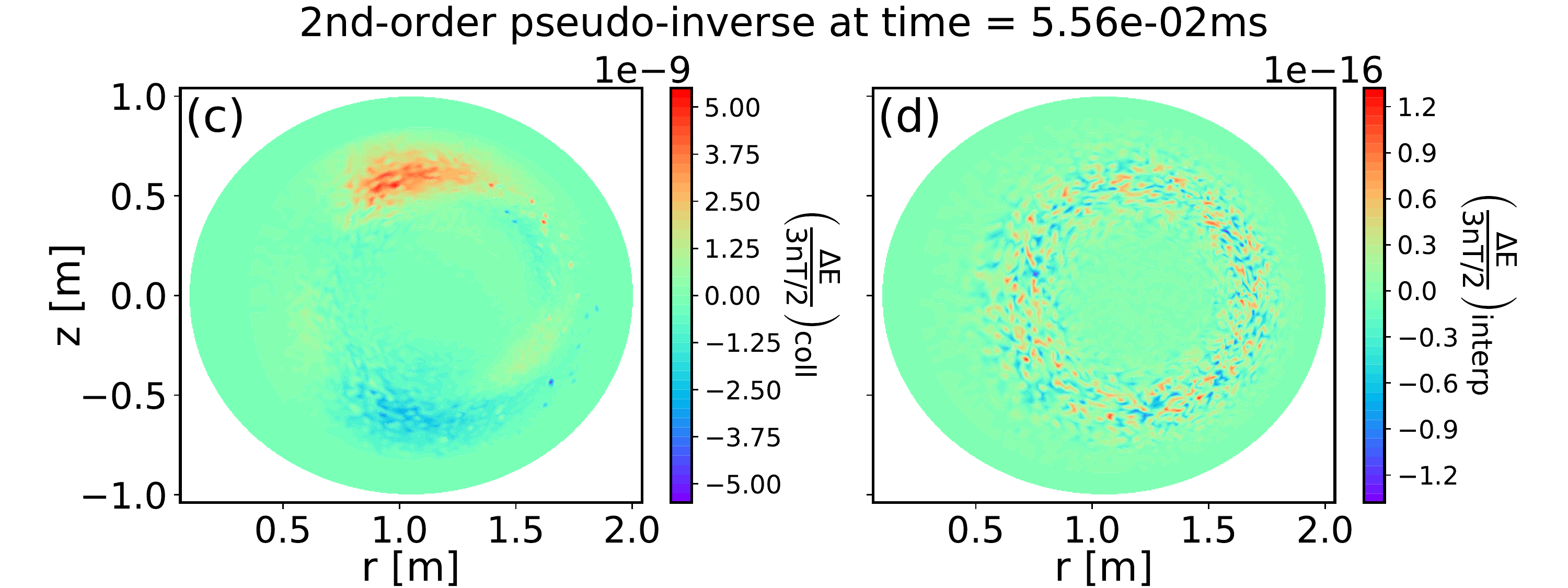}\\ \vspace*{0.2cm}
  \includegraphics[width=0.7\textwidth]{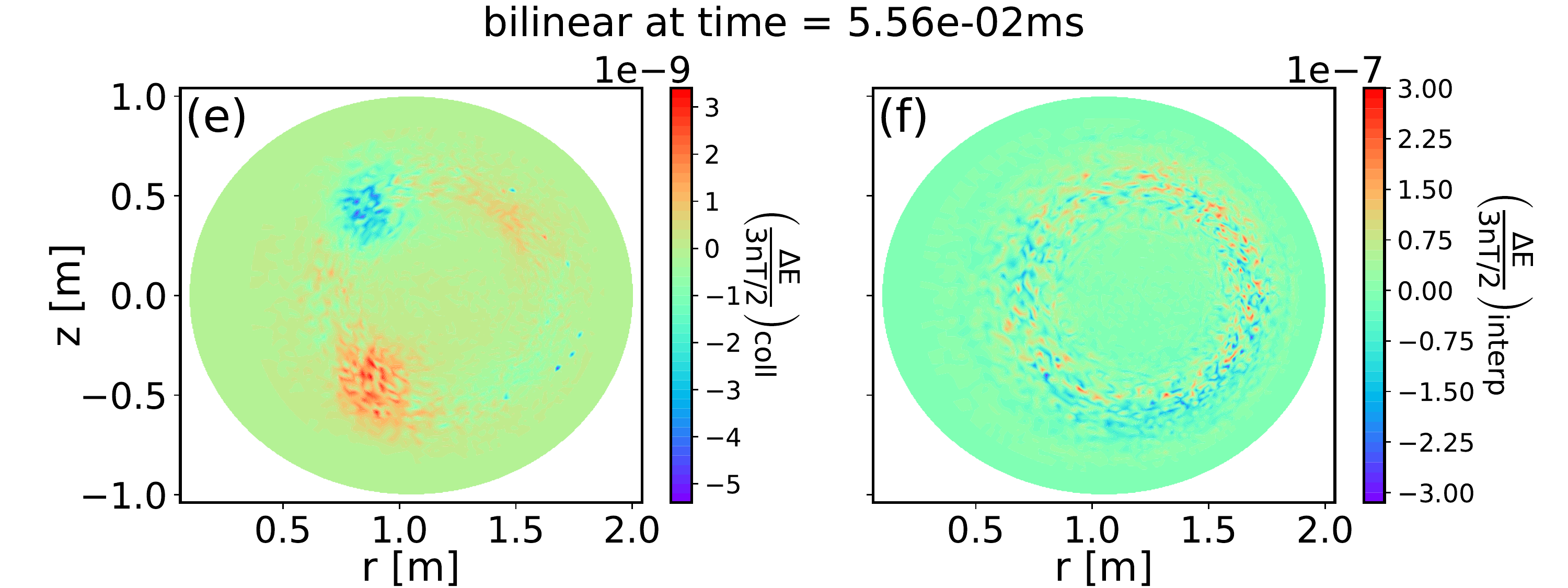}
\end{center}
\caption{
Color maps of the change in kinetic energy normalized to pressure $\displaystyle \frac{\Delta E}{3 n T / 2}$ (relative energy error) for the ions in all configuration space vertices of the poloidal cross-section in time step 200 of the simulation. The color maps show the change due to the collisions (a, c, e) and the change due to the interpolation (b, d, f), comparing the three different velocity mapping methods: $1^{\mathrm{st}}$-order ``pseudo-inverse'' (a, b), $2^{\mathrm{nd}}$-order ``pseudo-inverse'' (c, d) and ``bilinear'' (e, f).}
\label{fig:energyError}
\end{figure}

In Figs.~\ref{fig:densityError}-\ref{fig:energyError} we show the poloidal cross-section of the relative density error, the relative parallel momentum error and the relative kinetic energy error for the Coulomb collision operation and the velocity interpolation calculated in each configuration space vertex in a time step of the simulation. The relative density, momentum, and energy errors are normalized to background density $n$, momentum at thermal speed $\displaystyle m n v_{T} = n \sqrt{2 m T}$, and pressure $\displaystyle 3 n T / 2$, respectively. 

Fig.~\ref{fig:densityError} shows that all three velocity mapping methods exhibit a much smaller interpolation density error than collision density error. This implies that the change in density from the collision operation seen in the marker particle representation is unaffected by the velocity interpolation. 
Note that since the marker particle weights in the `$\mathrm{grid} \rightarrow \mathrm{particles}$' mapping are distributed differently depending on the method used, we cannot expect the collision error to look exactly the same when comparing between the different methods. What matters is that the error remains of the same size and the interpolation error is significantly smaller. 

For the momentum error in Fig.~\ref{fig:momentumError} we see that the ``pseudo-inverse'' methods have a much smaller interpolation error than the ``bilinear'' method. In fact, there are configuration space vertices where the ``bilinear'' method results in a larger interpolation error than the collision error. This is not ideal from a numerical point of view, even though the total error in the marker particle representation remains smaller than required by \xgc. 
The test case here has relatively benign properties with mild gradients and the interpolation error could be larger for realistic geometry and High-confinement mode profiles. However, it is possible to control the error by the number of marker particles in the simulation \citep{HagerJCP2016}.

As expected from theory, Fig.~\ref{fig:energyError} illustrates that the quadratic $2^{\mathrm{nd}}$-order ``pseudo-inverse'' method exhibits an interpolation error in kinetic energy that is negligible compared to the collision error. In contrast, for the two linear methods the interpolation error dominates over the collision error. In the figure we see that configuration space vertices located around mid-radius, where the pressure gradient is larger in the plasma (see Fig.~\ref{fig:PlasmaProfiles}), generally show a larger interpolation error. 

\begin{figure}
\begin{center}
  \includegraphics[width=0.9\textwidth]{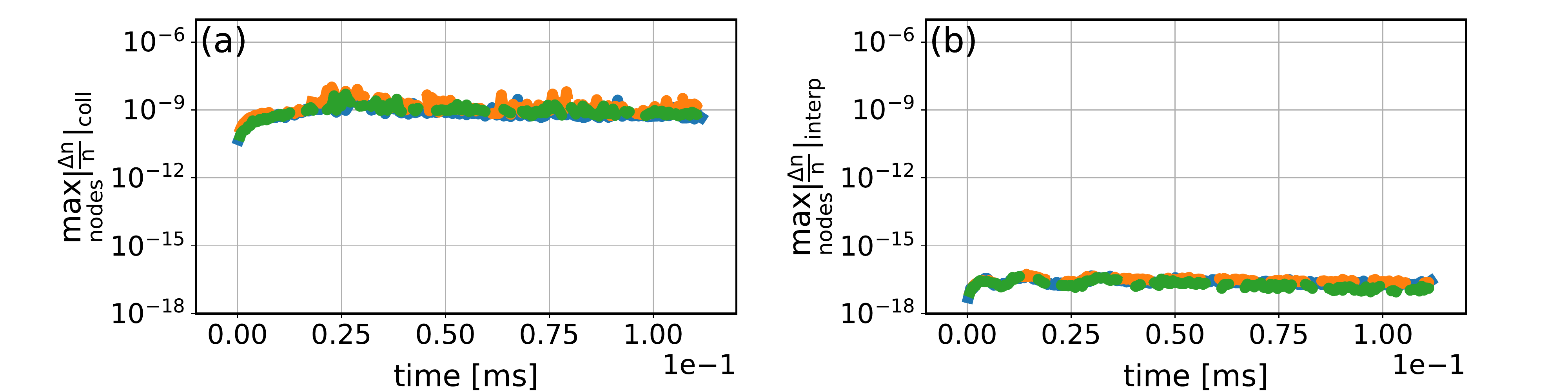}\\ \vspace*{0.2cm}
  \includegraphics[width=0.9\textwidth]{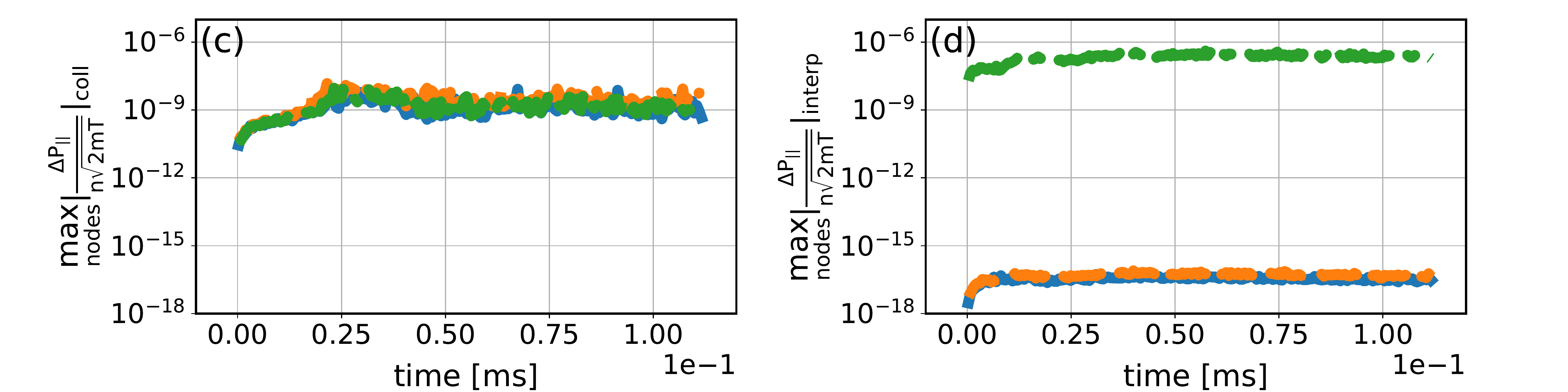}\\ \vspace*{0.2cm}
  \includegraphics[width=0.9\textwidth]{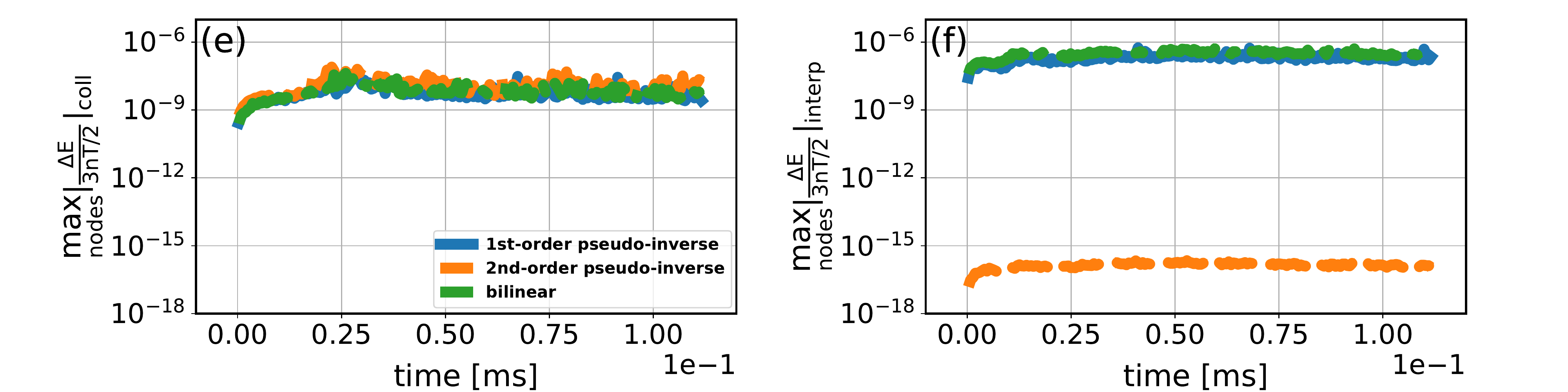}
\end{center}
\caption{
Time evolution of the maximum relative density error (a, b), maximum relative parallel momentum error (c, d) and maximum relative kinetic energy error (e, f) in any configuration space vertex (node) for the Coulomb collision operation (a, c, e) and the velocity interpolation (b, d, f), comparing the three different velocity mapping methods: $1^{\mathrm{st}}$-order ``pseudo-inverse'' (\textcolor{NewBlue}{\newfull}), $2^{\mathrm{nd}}$-order ``pseudo-inverse'' (\textcolor{NewOrange}{\longdashed}) and ``bilinear'' (\textcolor{NewGreen}{\dashdot}).}
\label{fig:TotalErrorEvolution}
\end{figure}

Figs.~\ref{fig:densityError}-\ref{fig:energyError} only illustrate the errors in a single time step. To demonstrate a consistently reduced interpolation error with the ``pseudo-inverse'' methods we show the time evolution of the maximum relative errors in any configuration space vertex in Fig.~\ref{fig:TotalErrorEvolution}. 
The collision errors in the left-hand figures, controlled by the tolerances in \xgc, are well below $10^{-7}$ throughout the simulation. The relative density error in the interpolation is consistently much smaller for all three methods, but the maximum momentum and energy errors are always larger than the corresponding collision errors for the ``bilinear'' mapping. The ``pseudo-inverse'' mappings perform as expected. The $1^{\mathrm{st}}$-order mapping results in a negligible momentum error in the interpolation but the energy error is of the same size as for the ``bilinear'' mapping. The $2^{\mathrm{nd}}$-order mapping shows negligible interpolation errors in both momentum and energy for all time steps. 
From Fig.~\ref{fig:TotalErrorEvolution} it is evident that the ``pseudo-inverse'' methods maintain the reduced interpolation errors throughout the simulation. 
We can thus conclude that the new velocity mapping implementation in \xgc~works as expected.

The energy error caused by the grid-to-particle bilinear interpolation in the core-region example shown here is not significant.  However, in the steep edge pedestal, this error could become significant. 
How the bilinear interpolation error could cause some inaccuracies in the edge pedestal physics and how the present quadratic pseudo-inverse interpolation can correct them will be the subject of a subsequent report.
%
%
\section{Discussion and conclusions}
\label{sec:conclusions}
We have updated the particle-in-cell code \xgc~with a new mass, momentum, and energy-conserving mapping between marker particles and the uniform rectangular grid in velocity space. This new technique, based upon the calculation of a pseudo-inverse, preserves moments up to the order of the finite element space used for the mapping. 
We have demonstrated that the interpolation error is several orders of magnitude smaller than the corresponding error in the Coulomb collision operation for a tokamak neoclassical test case, and a significant improvement compared to the default bilinear mapping used in \xgc. As a consequence, the energy conservation of the collision operation in the marker particle representation of the distribution function is improved. This is important for the fidelity of the edge plasma simulations where the plasma pressure slope can be extremely steep and the bilinear particle-grid mapping may not be sufficiently accurate. 
With the new technique, \xgc~is equipped with the capability to make both the collision and interpolation errors in energy arbitrarily small by selecting strict tolerances in the iterative solvers.

Our implementation is also a step towards integrating the Fokker-Planck-Landau collision operator with second order finite-element velocity grids and optional adaptive mesh refinement (AMR) in \xgc.
A fixed, rectangular velocity grid has the disadvantage of high-velocity cells being over-resolved, since fewer particles reside in these cells. The computational expense of the grid operations is therefore larger than necessary.
The application of AMR will be beneficial in multiple ways, e.g. by minimizing the need for particle resampling and the cost of continuum operations caused by over-resolved areas of the velocity grid, and by facilitating the inclusion of fast particle physics.


\section*{Acknowledgments}
%
This research was supported by the U.S. Department of Energy Office of Science ASCR and FES through SciDAC-4 Partnership Center for High-fidelity Boundary Plasma Simulation (HBPS) under the Contract No. DE-AC02-09CH11466 at Princeton University and at LBNL, and by the ECP program via funding to WDMApp (17-SC-20-SC). 
This research used resources of the National Energy Research Scientific Computing Center (NERSC), a U.S. Department of Energy Office of Science User Facility operated under Contract No. DE-AC02-05CH11231. 
Computing resources were also provided on the PPPL computer Traverse operated by the PPPL and Princeton Institute for Computational Science and Engineering (PICSciE). 
The authors are grateful to E.~F.~D'Azevedo, J.~Dominski and S.~Ku for fruitful discussions.



\bibliographystyle{jpp}

\bibliography{XGC_higher_order_velocity_interpolation}

\end{document}